\documentclass{elsart}
\usepackage{epsfig}

\begin{document}
\begin{frontmatter}
\title{One-Up On L1: Can X-rays Provide Longer Advanced
Warning of Solar Wind Flux Enhancements Than
Upstream Monitors?}

\author{M.R. Collier$^{a}$, T.E. Moore$^{a}$,
S.L. Snowden$^{b}$ and K.D. Kuntz${}^{c}$}



\address[GSFC]{GSFC/NASA Code 692, Greenbelt, MD. 20771}
\address[GSFC]{GSFC/NASA Code 662, Greenbelt, MD. 20771}
\address[USRA]{Astronomy Department, University of Maryland Baltimore County, 1000 Hilltop Circle, Baltimore, MD  21250}

\begin{abstract}
Observations of strong solar wind proton flux correlations with ROSAT X-ray rates along with high spectral resolution Chandra observations of X-rays from the dark Moon show that soft X-ray emission mirrors the behavior of the
solar wind. In this paper, based on an analysis of an X-ray event observed by XMM-Newton resulting from charge exchange of high charge state solar wind ions and contemporaneous neutral solar wind data, we argue that X-ray observations may be able to provide reliable advance warning, perhaps by as much as half a day, of dramatic increases in solar wind flux at Earth. Like neutral atom imaging, this provides the capability to monitor the solar wind remotely rather than in-situ.  
\end{abstract}

\end{frontmatter}

\section{Introduction}
The space physics community relies heavily on observatories perched
at the L1 point, about 235~R${}_{\rm E}$ upstream, to provide a profile
of the plasma and field structure of the outward convected solar wind and 
to serve as sentries for potentially dangerous geoeffective conditions.
The first goal is compromised by the fact that the solar wind is a 
highly-structured medium [Collier et al., 2000] 
both in plasma [e.g. Richardson and Paularena, 2001] and in 
magnetic field [e.g. Collier et al., 1998; Crooker et al., 1982]
which, especially for observations far off the Sun-Earth line, 
can limit the usefulness of these upstream monitors.
The second goal is stymied by the fact that observations at L1 provide
only about an hour, frequently considerably less time, to react to
potential threats.

Because of these limitations, it behooves the community to at least consider
alternative methods of monitoring the interplanetary medium. One
of these methods employs neutral atom imaging, which samples, among
other products, solar wind charge exchange with exospheric neutral
atoms surrounding the Earth [e.g. Collier et al., 2001]. This technique
has the advantage of being able to infer the solar wind proton flux close
to the Earth without actually being outside the Earth's magnetosphere.

However, solar wind charge exchange produces not only neutral atoms, but
also soft X-ray photons [Cravens, 1997; Cravens et al., 2001; Wargelin et al., 2004; Gunell et al., 2004]. These soft X-rays associated
with the solar wind charge exchange process result from high charge state
solar wind ions, for example O${}^{+7}$ and O${}^{+8}$, charge exchanging
with neutrals in the interplanetary medium, generally either interstellar
or exospheric
neutrals, and ending up in an excited state. When the excited state relaxes,
soft X-rays are emitted.

\begin{figure}
\begin{center}
\epsfig{file=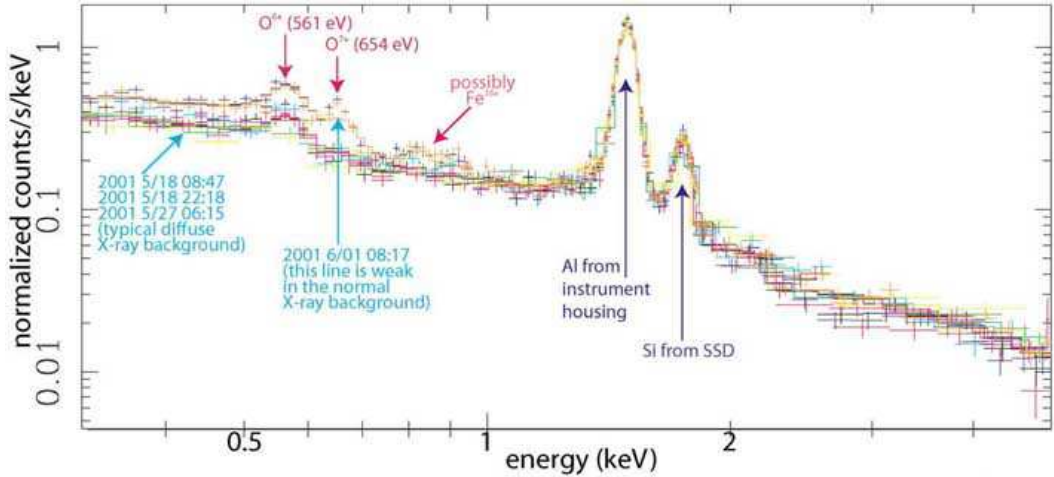,height=2.5in,width=5.5in}
\end{center}
\caption{Spectra from four Hubble Deep Field North observations taken by
XMM-Newton during late May and early June 2001. The solar wind charge
exchange (SWCX) spectrum consists of emission lines only and results  primarily
from charge exchange with interstellar or exospheric hydrogen.}
\end{figure}

\begin{figure}
\begin{center}
\epsfig{file=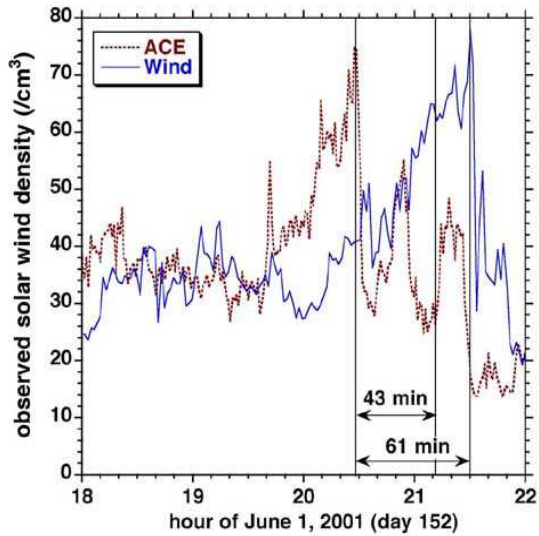,height=3.0in,width=3.0in}
\end{center}
\caption{ACE and Wind solar wind densities during the time of the
XMM-Newton observations. ACE is farther upstream than Wind and so sees the solar
wind structure before Wind. However, Wind observes the structure lagged by over an hour rather than 43 minutes, the solar wind convection time between the two spacecraft, indicating tilted solar wind phase fronts.}
\end{figure}

\begin{figure}
\begin{center}
\epsfig{file=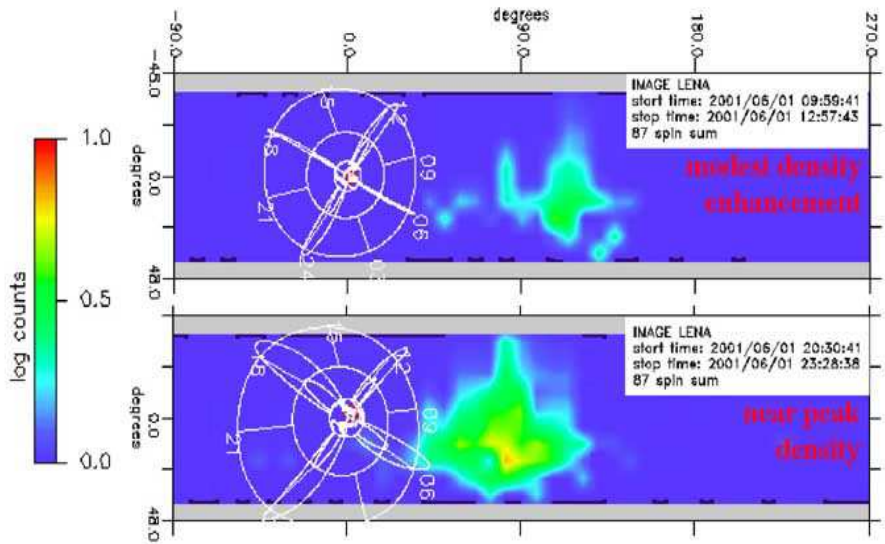,height=3.03in,width=4.88in}
\end{center}
\caption{Two LENA atomic hydrogen images from a period before the large density enhancement (top) and during the density plug (bottom). The Earth is at the origin and the direction closest to the Sun is near 90${}^\circ$ on the horizontal axis. During the density enhancement IMAGE/LENA observed sun pulse brightening resulting from enhanced SWCX.}
\end{figure}

Figure~1 shows spectra from four XMM-Newton observations in the direction 
of the Hubble Deep Field (HDF) North. 
Three of these observations are statistically
consistent with each other and with the typical diffuse background. 
Much of the
fourth observation, however, shows an elevated low energy component
with clear lines and a spectrum that is consistent with that 
expected from charge exchange emission between the highly ionized
solar wind and interstellar neutrals [Snowden et al., 2004].
This observation is concurrent with an enhancement in the solar wind flux
measured by Wind and ACE, as shown in Figure~2. 
The solar wind charge exchange spectrum consists of emission lines only
(no detected continuum component) and the  primary neutral participant in the charge exchange is either interstellar or exospheric hydrogen.

Additionally, observations from LENA on IMAGE show evidence for enhanced neutral solar wind, a product, along with soft X-rays, of increased solar wind charge exchange. The two panels in Figure~3 show two LENA atomic hydrogen images summed over 87 spins or close to three hours which cover a bit over half the sky. The
Earth is located at the origin with a nominal auroral oval in red. Dipole magnetic field lines corresponding to L=3 and L=6.6 along with the magnetic local time are indicated on the plot. The color bar shows, logarithmically, 
the total number of counts. The top panel is prior to the passage of the density enhancement with IMAGE at (2.9, 1.0, 7.5)~R${}_{\rm E}$ in geocentric solar ecliptic coordinates at 1130~UT. The dark green signal near 90${}^\circ$
on the horizontal axis is the spin direction closest to the Sun. The bottom
panel is near the period of maximum solar wind density with IMAGE at
(0.6, 3.0, 5.5)~R${}_{\rm E}$ in geocentric solar ecliptic coordinates at 2200~UT. The bright yellow signal near the Sun direction shows that the neutral hydrogen flux coming from the direction of the Sun has increased by about a factor of two between the two orbits. Figure~4 shows that the ten minute summed hydrogen count rate from the Sun direction during this period tracks the ACE ionized solar wind flux when the data are shifted by 1.28 hours to account for convection time from L1 to Earth.

\begin{figure}
\begin{center}
\epsfig{file=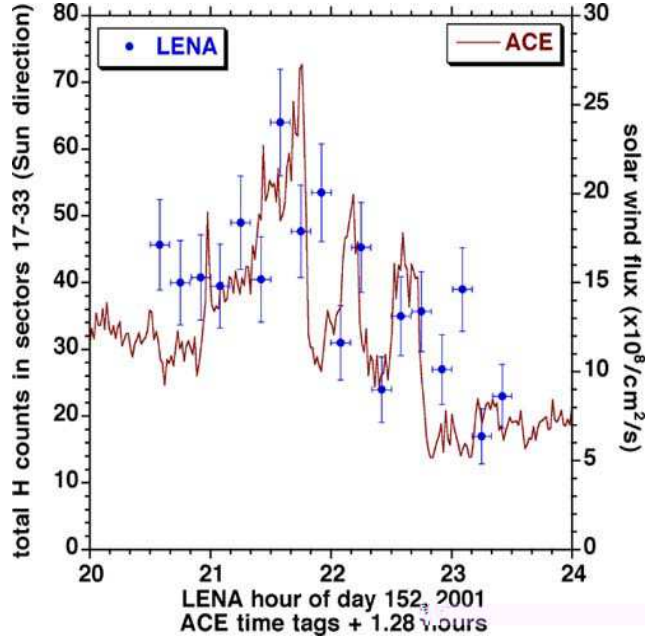,height=3.3in,width=3.3in}
\end{center}
\caption{IMAGE/LENA neutral hydrogen count rate in the direction of the Sun during the time period shown in the lower panel of Fig.~3 along with the ACE ionized solar wind flux shifted by 1.28 hours. The LENA neutral solar wind data
track the ACE solar wind flux reasonably well during this period.}
\end{figure}

\section{Light Curve Mystery}

Snowden et al. [2004], however, noted that the SWCX emission observed
in the X-ray light curve did not exhibit the same variability that the local
solar wind flux measured by ACE did. As shown in Figure~5, 
over the six hour period prior
to the peak flux passing the Earth, the light curve (blue)
remained essentially
constant, while the solar wind flux (red) increased
by a factor of six. Yet, the X-ray emission fell off concurrently with 
the end of the solar wind enhancement.
The mystery is why the light curve was elevated for about eleven hours
prior to the passage of the solar wind flux enhancement. If the solar
wind charge exchange occurs locally, for example with exospheric neutrals
in the magnetosheath [Robertson and Cravens, 2003], then it should rise
as well as fall with the solar wind flux. If it charge exchanges far away,
then the drop off might not be expected to occur right as the solar wind
structure passes the Earth.

\begin{figure}
\begin{center}
\epsfig{file=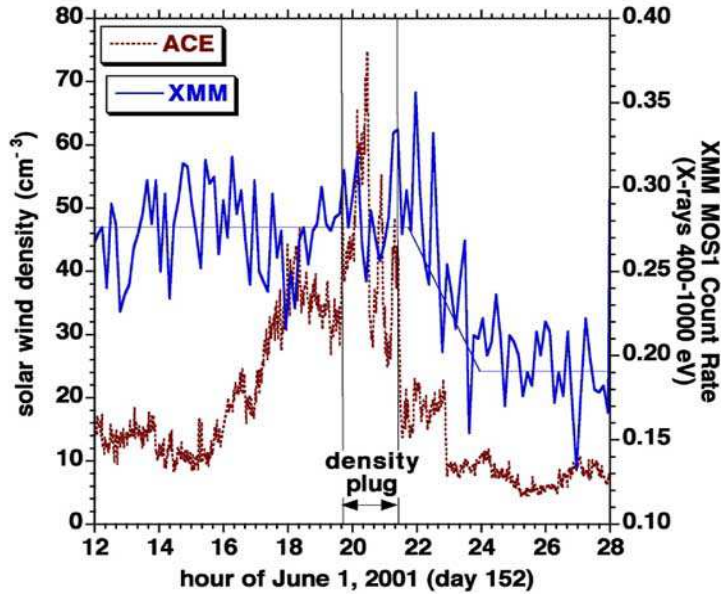,height=3.3in,width=3.96in}
\end{center}
\caption{Comparison between the ACE solar wind density and the XMM soft
X-ray rate on June 1, 2001 (day 152). Note that the XMM count rate is relatively steady prior to the time when the ACE density plug passes the Earth, but falls off over about a two hour period, the approximate width of the solar wind density enhancement.}
\end{figure}

Determining the onset time of the X-ray flux enhancement and its relationship to the solar wind data would likely resolve the mystery. Unfortunately, although Fig.~5 begins at 1200~UT, the X-ray flux was already enhanced from the beginning of the observation period at 0817~UT until its drop-off a bit
before 2200~UT. There are also some spikes in the first and last couple of
hours of the observations due to local contamination from soft protons 
(soft-proton flaring).

\section{The Structured Interplanetary Medium}

In an attempt to try to determine the source of the unexpected behavior of
the light curve, we used the Wind and ACE solar wind plasma data and the location of these two spacecraft at the time of the XMM-Newton observations
to determine the orientation of the phase fronts in the solar wind. 
Structures in the plasma and fields in the interplanetary medium over length
scales of many tens of Earth radii are approximately planar and can be propagated as such [e.g. Collier et al., 1998; Weimer et al. 2003].

\begin{figure}
\begin{center}
\epsfig{file=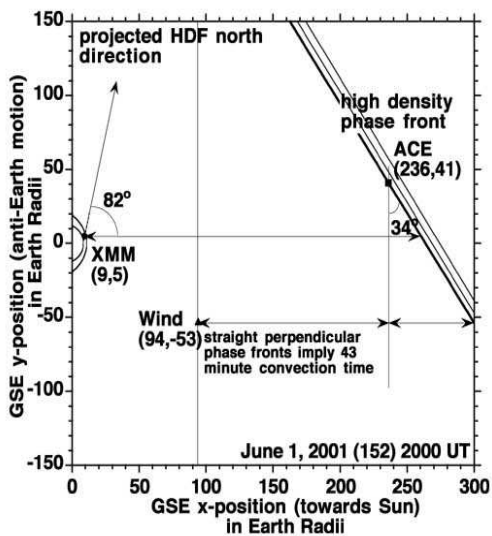,height=3.0in,width=3.0in}
\end{center}
\caption{The spacecraft positions and solar wind phase fronts at the time of the XMM-Newton observations. The high density plug in the  solar wind is oriented
34${}^\circ$ with respect to the Sun-Earth axis. Consequently, the XMM line-of-sight intersects the high density plane until after it passes the Earth.}
\end{figure}

Figure~6 shows the Wind and ACE spacecraft positions relative to the Earth at this time. From 2000-2200~UT on June 1, 2001, the solar wind convection speed
was about 351~km/s, implying that if the solar wind density phase fronts were
oriented perpendicular to the Sun-Earth line, there would be a 43~minute lag
between the data observed at ACE and that observed at Wind. 

However, as shown in Fig.~2, the observed lag time is about 61~minutes, based
on a time lagged cross correlation analysis [e.g. Collier et al., 1998],
which matches similar profiles in the plasma data. This
implies that the solar wind plasma phase fronts are tilted at an angle of about
34 degrees with respect to the Sun-Earth line based on the locations of ACE and
Wind. Consequently, the line-of-sight of XMM-Newton,
which is shown on Fig.~6 along with the projected Hubble Deep Field North
direction at about
82 degrees with respect to the Sun-Earth line, intersects the planar solar wind
density enhancement. If the orientation of the phase front had been at a sizable negative angle, then the XMM-Newton line-of-sight would not have intersected it,
and no enhancement would have been observed prior to the density
structure's passage. The large scale size of these structures and planar geometry mean that the light curve,
if the structure intersects XMM-Newton's line-of-sight, remains constant until
the structure passes the spacecraft. 
The steady fall-off of the X-ray intensity over an approximately two hour period
may represent the amount of time it takes the density plug indicated in Fig.~5
to pass the Earth.
Note, however, that the direction of the phase fronts in this case is very close to the nominal spiral angle of the
interplanetary magnetic field at 1~AU, so that such an orientation will be more
common than one with a phase front at a sizeable negative angle [Richardson and Paularena, 2001].

\section{Conclusions}
We have discussed an event reported by Snowden et al. [2004] observed by
XMM-Newton in which solar wind charge exchange is clearly occurring. However,
oddly enough, the light curve associated with this event was elevated until the
time at which the solar wind flux enhancement passed the Earth. We show that
the behavior of 
the light curve may be explained based on the phase front orientation of the
solar wind density enhancement. Because the X-ray intensity was elevated
for about 11 hours prior to the passage of the solar wind structure,
this suggests the possibility that X-ray imaging
may be able to provide longer advanced warning of solar wind flux enhancements
than upstream monitors at L1.


\end{document}